# An Efficient Encryption Algorithm for P2P Networks Robust Against Man-in-the-Middle Adversary

Roohallah Rastaghi

**Department of Electrical Engineering, Aeronautical University of Since & Technology,**
Tehran, Iran

**Abstract**

Peer-to-peer (P2P) networks have become popular as a new paradigm for information exchange and are being used in many applications such as file sharing, distributed computing, video conference, VoIP, radio and TV broadcasting. This popularity comes with security implications and vulnerabilities that need to be addressed. Especially, due to direct communication between two end nodes in P2P networks, these networks are potentially vulnerable to "Man-in-the-Middle" attacks.

In this paper, we propose a new public-key cryptosystem for P2P networks that is robust against Man-in-the-Middle adversary. This cryptosystem is based on RSA and knapsack problems. Our precoding-based algorithm uses knapsack problem to perform a permutation and pads random data to the message. We show that comparing to other proposed cryptosystems, our algorithm is more efficient and it is fully secure against an active adversary.

## 1. Introduction

The popularity of information technologies is increasing significantly. Increased usage of these technologies will cause to a huge increment in the number of users. This results in lack of sufficient servers in client-server configurations in spite of having a large enough bandwidth. Moreover, the data that is needed by the client may not be available on the servers that are connected to it. To address this problem, an alternative to traditional client-server configurations is introduced, which is called Peer-to-Peer (P2P) network. A P2P network is a group of computer nodes, which construct their own open unrestricted sharing networks on top of the Internet architecture. P2P technology is widely used in file sharing protocols such as BitTorrent and Dropbox, as well as in instance message communication systems such as Skype.

The P2P system has reached to the fourth generation. This generation applies decentralized servers in the network. In fact each device is classified as a peer and simultaneously it may be a client or server peer. The fourth generation also supports streams over P2P networks. The success of P2P networks was not limited to file sharing, video conference, distributed computing, VoIP, radio and TV broadcasts are other applications that became popular and presented the fourth generation P2P networks. Podcasts is one of the most famous types.

Due to the decentralized and peer-relying nature of P2P networks, they are susceptible to many general attacks. Similar to traditional Internet, P2P networks are vulnerable to attacks, such as Denial-of-Service (DoS) attack, Distributed Denial-of-Service (DDoS) attack, Worm and virus propagation, Pollution attack and Man-in-the-middle attack. P2P networks can also be the victims of some P2P specific attacks. Rational attack, Index Poisoning attack, Sybil attack, and Eclipse attack are P2P specific attacks. To protect these general attacks, technologies and mechanisms for ensuring network safety (for example [7, 11]) and the common network knowledge, such as encryption mechanisms and authentication technologies. Also, some well-known safety measures, such as firewalls, anti-virus software, and secure operating systems, provide the relative defensive strategies. The Man-in-the-middle attack is one the most pernicious attacks against P2P networks and may reduce the performance of these networks significantly [9, 14]. Simulation results in [9] show that this type of attack is a serious problem, and using only a small number of malicious nodes, it is possible to corrupt a very large number of requests. Man-in-the-middle attack is an indirect intrusion, where the attacker inserts its node undetected between two nodes. It is a security threat in which such peers get between the receiving peer and the sending peer in a P2P network and sniffs the information being sent. It is typically used to read a public-key encrypted conversation. Gnutella, which is one of the largest P2P networks, is highly vulnerable to these kinds of attacks. The most common instance is modification of a Query Hit message some malicious node in the path from sender to receiver in Gnutella. The modified Query Hit directs the downloading request to a non-existent node or an unreliable or a malicious node. To best of our knowledge, there exist no direct and effective defense strategies against this type of attack.

In this paper, we propose an efficient cryptographic algorithm that is robust against active man-in-the-middle attacks, in which —in addition to eavesdropping— the adversary is able to insert, delete, or arbitrarily modify messages sent from one user to another. Few techniques exist for dealing with active attacks, and designing practical protocols secure against such attacks remains a key challenge. We provide new, provably-secure protocols that prevent such attacks. Proofs of security are in the standard cryptographic model and rely on knapsack problem assumptions.

The rest of this paper is organized as follows: In the following section, we briefly explain some mathematical background and definitions. Then, in Section 3, we introduce our proposed cryptosystem. Performance and security analysis of this cryptosystem will be discussed in Section 4.

## 2. Preliminary

In this section, we briefly provide some mathematical background and security concepts and notions that are required to comprehend the paper.

2.1 Mathematical Background

The first knapsack-type public key cryptosystem (PKC) was introduced by Merkle and Hellman [13]. Since its proposal, knapsack-type PKC have been widely studied and many knapsack-based PKCs were developed. Nevertheless, almost all knapsack cryptosystems were shown insecure against some known attacks, such as low density attack [1, 3], Shamir's attack [12], lattice basis reduction algorithms [8], etc.

**Definition1** (Subset sum problem (SSP)). *A set of positive integers* $(a_1, a_2, \ldots, a_n)$ *and positive integer $s$ is given. Whether there is a subset of the $a_i$'s that their sum equal to $s$. That is equivalent to determine whether there are variables $x_i \in \{0,1\}, 1 \leq i \leq n$ such that*

$$\sum_{i=1}^{n} x_i . a_i = s$$

The Subset sum problem is a particular case of the 0-1 knapsack problem. The subset sum problem has been proven to be *NP-complete*. The computational version of the subset sum problem is *NP-hard* [5]**.**

**Definition2** (Super-increasing sequence). *The sequence* $(a_1, \ldots, a_n)$ *of positive integers is a super-increasing sequence, if* $a_i > \sum_{j=1}^{i-1} a_j$ *for all* $i \geq 2$.

There is an efficient greedy algorithm to solve the subset sum problem if the $a_i$'s area super-increasing sequence: Just subtract the largest possible value from $s$ and repeat.

Knapsack public-key encryption schemes are based on the subset sum problem, which is NP-complete [4]. The basic idea is to select an instance of the subset sum problem that is easy to solve, and then to disguise it as an instance of the general subset sum problem, which is hopefully difficult to solve. The original knapsack set can serve as the private key, while the transformed knapsack set serves as the public key.

**Definition3** (RSA problem [10]). *Given an RSA public key* $(n, e)$, *where* $n = p.q$ *and* $p, q, (p-1)/2, (q-1)/2$ *are large prime numbers, $e$ is an odd integer such that* $\gcd(e, \varphi(n)) = 1$ *and* $b \in_R \mathbb{Z}_n^*$. *Finding* $a \in \mathbb{Z}_n^*$ *such that* $a^e = b \bmod n$ *is referred as the RSA problem.*

2.2 Man-in-the-middle attack and IND-CCA2 Security

The *semantic security* (a.k.a. indistinguishability) against adaptive chosen ciphertext attacks (IND-CCA2) is the strongest known notion of security for the public key encryption schemes. This notion of security simply means that a cryptogram should not reveal any useful information about the message. This type of attack is the most powerful attack, which is defined by Rackoff and Simon [6]. In this scenario, the adversary has seen the "target" cryptogram before having access to the decryption oracle. The adversary is not allowed to ask the decryption of the "target" cryptogram, but can obtain the decryption of any relevant cryptogram (even modified ones based on the target cryptogram). A cryptosystem is said to be IND-CCA2 secure if the cryptanalyst fails to obtain any partial information about the plaintext relevant to the "target" cryptogram. This notion is known to suffice for many applications of encryption in the presence of active attackers —a man-in-the-middle— including secure P2P transition, secure communication, auctions, voting schemes, and many others. Indeed, CCA security is commonly accepted as the security notion of choice for encryption schemes that are to be "plugged in" to a protocol running in an arbitrary setting. So, if an encryption algorithm in the P2P networks is secure against active adversary, then the malicious nodes cannot eavesdrop, delete, insert or arbitrarily modify messages sent from one user to another. Man-in-the-middle Attack is also called Parallel Sessions attack, because it requires two sessions to be running in parallel. The intruder should exchange messages with the first and the second victims, claiming to be the other party in each connection.

In recent year, attempt for designing cryptosystem secure against adaptive chosen ciphertext is increased, but no efficient and secure cryptosystem in this area is introduced.

**Definition 5** (Public-key encryption). *A public-key encryption scheme PKE is a triple of probabilistic polynomial time (PPT) algorithms $(\mathcal{G}, \mathcal{E}, \mathcal{D})$ such that:*

- *The randomized key generation algorithm $\mathcal{G}$ takes as input a security parameter $1^k$ and outputs a public key $pk$ and a secret key $sk$. We write $(pk, sk) \leftarrow \mathcal{G}(1^k)$.*
- *The randomized encryption algorithm $\mathcal{E}$ takes as input a public key $pk$ and a message $m \in \{0,1\}^*$, and outputs a ciphertext $C$. We write $C \leftarrow \mathcal{E}_{pk}(m)$.*
- *The decryption algorithm $\mathcal{D}$ takes as input a ciphertext $C$ and a secret key $sk$. It returns a message $m \in \{0,1\}^*$ or the distinguished symbol $\perp$. We write $m \leftarrow \mathcal{D}_{sk}(C)$.*

*We require that for all $(pk, sk)$ outputs by $\mathcal{E}$, all $m \in \{0,1\}^*$, and all $C$ outputs by $\mathcal{E}_{pk}(m)$ we have $\mathcal{D}_{sk}(C) = m$.*

**Definition 6** (CCA security [2]). *A public-key encryption scheme PKE is secure against adaptive chosen-ciphertext attacks (i.e. IND-CCA2) if the advantage of any PPT adversary $\mathcal{A} = (\mathcal{A}_1, \mathcal{A}_2)$ in the following game is negligible in the security parameter $k$:*

$$(pk, sk) \xleftarrow{\$} \mathcal{G}(1^k)$$
$$(m_0, m_1) \xleftarrow{\$} \mathcal{A}_1^{\mathcal{D}}(pk)$$
$$b \xleftarrow{R} \{0,1\}$$
$$C^* \xleftarrow{\$} \mathcal{E}(pk, m_b)$$
$$b' \xleftarrow{\$} \mathcal{A}_2^{\mathcal{D}}(C^*)$$

*The attacker may query a decryption oracle with a ciphertext $C$ at any point during its execution, with the exception that $\mathcal{A}_2$ may not query the decryption oracle on $C^*$. The decryption oracle returns $m \leftarrow \mathcal{D}_{sk}(C)$. The attacker wins the game if $b = b'$ and the probability of this event defined as $\Pr_{\mathcal{A}}[\text{succ}]$. An attacker's advantage is defined to be*

$$\text{Adv}_{\mathcal{A}}^{\text{Ind-CCA2}}(k) = |\Pr[b = b'] - 1/2| < \varepsilon. \quad (1)$$

## 3. The Proposed Cryptosystem

In this section, a new cryptosystem secure against man-in-the-middle attack will be proposed. This cryptosystem is based on RSA and Knapsack problems. Our scheme is a precoding-based algorithm that uses knapsack problem for performing permutation and random data padding to the message.

### 3.1 Key generation algorithm

1. Randomly choose $k$ odd positive integers $c_1, \ldots, c_k$, where $c_k = 1$, and $c_i$ has $(l-i)$-bits binary length for $i = 1, \ldots, l-i$.
2. Compute $b_i = 2^{i-1} \cdot c_i$ for $i = 1, \ldots, k$.
3. Randomly choose a modulus $M > \sum_{i=1}^{k} b_i$ and a multiplier $w$ with $0 < w < M$ and $\gcd(M, w) = 1$.
4. Compute $a_i = b_i \cdot w \mod M$ for $i = 1, \ldots, k$.
5. Generate two large random and distinct prime numbers $p, q$ such that $(p-1)/2$, $(q-1)/2$ be also large prime numbers. Compute $n = p \cdot q$ and $\varphi(n) = (p-1) \cdot (q-1)$.
6. Randomly choose an integer $e$ such that $1 < e < \varphi(n)$ and $\gcd(\varphi(n), e) = 1$.
7. Use the extended Euclidean algorithm to compute the unique integer $d$, $1 < d < \varphi(n)$ such that $e \cdot d = 1 \mod \varphi(n)$.

$\{(a_1, \ldots, a_k), n, e\}$ is the private key and $\{(b_1, \ldots, b_k), p, q, w, M\}$ is the public key of the cryptosystem.

### 3.2 Encryption algorithm

To encrypt message $m \in \{0,1\}^k$ with $k < n$, the sender executes the following steps:

1. Generate pseudorandom string $X = (x_1, \ldots, x_k)$ with hamming weight $h$ such that $v = k/h$ be an integer.
2. Computes $\tilde{m} = m \oplus X$ and divide $m'$ into $h$ blocks with equal length $v$
$$m' = (d_1 \| \ldots \| d_h).$$
3. Using pseudorandom string $X = (x_1, \ldots, x_k)$, performs a random permutation on the message blocks $d_i, 1 \leq i \leq h$ and pad some confuse data blocks to them such that:

$$d_i' = \begin{cases} d_{\sum_{j=1}^{i} x_j} & \text{if } x_i = 1 \\ \text{confuse data} & \text{if } x_i = 0 \end{cases}, 1 \leq i \leq k.$$

In this case, $k-h$ confuse data blocks with equal length $s = \lfloor (n-k)/(k-h) \rfloor$ are pad to the message blocks $d_i, 1 \leq i \leq h$, to produce the processed message $m' = (d_1' \| \ldots \| d_k')$.

4. Suppose $y$ be the corresponding decimal representation of $m'$. Compute:

$$c_1 = y^e \bmod n, \quad c_2 = \sum_{i=1}^{k} a_i . x_i$$

and send ciphertext $c = (c_1, c_2)$ to the receiver.

## 3.3 Decryption algorithm

Receiver after receiving $c = (c_1, c_2)$, does the following steps to recover the plaintext $m$ from ciphertext $c = (c_1, c_2)$:

1. Compute $y = c_1^d \bmod n$
2. Compute vector $X$ as

$$r = c_2 . w^{-1} \bmod M = \sum_{i=1}^{k} x_i . b_i,$$

$$x_1 = r \bmod 2, \quad x_i = \frac{r - \sum_{i=1}^{i-1} x_i . b_i}{2^{i-1}} \bmod 2, \quad 2 \leq i \leq k$$

Note that $M > \sum_{i=1}^{k} b_i$. Thus the vector $X = (x_1, \ldots, x_k)$ that demonstrates position and length of data (and confuses data) blocks is calculated. Compute $h = \sum_{i=1}^{k} x_i$ and $s = \lfloor (n-k)/(k-h) \rfloor$.

3. The length of data blocks is equal to $v = (|m'| - (k-h) \times s)/h$ where $|m'|$ is the binary length of $y$. So, the length of data (and confuses data) blocks and position of them are explicit, therefore the receiver simply can separate confuse data from processed message $m'$ and recover $\tilde{m}$.
4. Outputs $m = \tilde{m} \oplus X$.

## 4. Performance and Security Analysis

The performance-related issues can be discussed with respect to the computational complexity of key generation, key sizes, encryption and decryption speed, and information rate.

The proposed cryptosystem features fast encryption and decryption. The encryption only carries out one exponentiation, an encoding, $\mathcal{O}(k)$ XOR and additions.

The decryption roughly needs one modular multiplication and one modular exponentiation. Altogether, the efficiency of our scheme is comparable to the classical RSA cryptosystem, while our construction is IND-CCA2.

### 4.1 Security Analysis of proposed cryptosystem

The security of our proposed cryptosystem is examined by considering known and potential attacks. The RSA cryptosystem is a well-studied encryption algorithm in the cryptology community. Security of this cryptosystem is based on integer-factoring problem. To date, there doesn't exist a polynomial time algorithm for solving integer-factoring problem or solving RSA problem. The proposed knapsack problem is secure under known attacks, such as Low-density attack, Shamir's attack, and lattice-based reduction algorithms [15].

**Theorem1:** *The proposed cryptosystem is secure against adaptive chosen ciphertext attack (i.e. IND-CCA2) in the standard model.*

**Proof:** Suppose that $\mathcal{A}$ is a PPT algorithm against IND-CCA2 security of the proposed cryptosystem. Let $c^* = (c_1^*, c_2^*)$ be the challenge ciphertext corresponding to the challenge message $m_b$, $b \in \{0,1\}$, where $c_1^* = \mathcal{E}_{RSA<n,e>}(y)$ and $c_2^* = \sum_{i=1}^{k} a_i . x_i$. Since a decryption query on the challenge ciphertext is forbidden by the CCA2-experiment, so if $c_1 = c_1^*$, then $c_2 \neq c_2^*$ and vice versa. We consider two potential cases $c = (c_1, c_2^*)$ and $c = (c_1^*, c_2)$ that may be queried to decryption oracle by the adversary $\mathcal{A}$. Note that queried ciphertext of the form $c = (c_1 \neq c_1^*, c_2 \neq c_2^*)$ does not reveal any useful information about challenge bit.

**Case1.** In this case, the CCA2 adversary queries on the ciphertexts of the form $c = (c_1, c_2^*)$. For this case, the decryption oracle computes $X = \mathcal{D}_{Knap}(c_2) = \mathcal{D}_{Knap}(c_2^*) = X^*$, and since $c_1 \neq c_1^*$, we have $y = \mathcal{D}_{RSA<n,e>}(c_1) \neq \mathcal{D}_{RSA<n,e>}(c_1^*) = y^*$. We should consider two possible cases:

1) $\tilde{m} = \tilde{m}^*$ while $y \neq y^*$. In this case the decryption oracle outputs $m_b = \tilde{m}^* \oplus X^*$ and the CCA2 adversary wins the game. The probability that this event occur is $2^{-(n-k)}$ that depends on the values of $n$ and $k$. For example, for

$n = 2k$ the success probability is $2^{-k}$, which is negligible (note that in the decryption algorithm we assume $n > k$).

2) $\tilde{m} \neq \tilde{m}^*$ and $y \neq y^*$. For such cases the decryption oracle outputs random string $m = \tilde{m} \oplus X^*$. Hence, challenge bit is information theoretically hidden to CCA2 adversary and his advantage to guessing challenge bit is 0.

Hence, the overall advantage of the CCA2 adversary in **Case1** is negligible.

**Case2.** In this case, the CCA2 adversary queries on the ciphertexts of the form $c = (c_1^*, c_2)$. For this case, the decryption oracle computes $X = \mathcal{D}_{\text{Knap}}(c_2) \neq \mathcal{D}_{\text{Knap}}(c_2^*) = X^*$, $y = y^*$, and then pick up message blocks from $y^*$ based on the bits of random string $X = (x_1, \ldots, x_k)$. So, $\tilde{m} \neq \tilde{m}^*$ and the decryption oracle outputs random string $m = \tilde{m} \oplus X$. Therefore, challenge bit is information theoretically hidden to CCA2 adversary and his advantage to guess challenge bit is 0.

From **Case1** and **Case2**, the overall advantage of the CCA2 adversary, i.e. $\text{Adv}_{\mathcal{A}}^{\text{Ind-CCA2}}(k)$, to guess challenge bit is negligible, which finishes the proof.

**Remark1.** *In* **Case2**, *the adversary $\mathcal{A}$ can modify ciphertext $c_2$ to guess challenge bit. The optimum choice for modifying $c_2$ such that it becomes a legitimate ciphertext is that the attacker randomly chooses $a_i$ from public weight $(a_1, \ldots, a_k)$ and produce ciphertext $c_2 = c_2^* + a_i$ and then queries it to the decryption oracle. If $a_i \in c_2$, then ciphertext $c_2$ is not a legitimate ciphertext and the knapsack cryptosystem outputs $\bot$ (reject). If $a_i \notin c_2$, then $c_2 = c_2^* + a_i$ is a legitimate ciphertext and the knapsack cryptosystem outputs plaintext $X = (x_1^*, \ldots, x_i = 1, \ldots, x_k^*)$ which is differ from $X^* = (x_1^*, \ldots, x_i^* = 0, \ldots, x_k^*)$ in the i-th bit. In such case we have $h = h^* + 1$ and picked up message blocks from $y^*$ based on the bits of $X = (x_1^*, \ldots, x_i = 1, \ldots, x_k^*)$ contains the bits of challenge message. Thus, the CCA2 adversary can guess challenge bit from $m = \tilde{m} \oplus X^*$ (the output of the decryption algorithm).*

*The probability that $c_2 = c_2^* + a_i$ becomes a legitimate ciphertext is $2^{-(k-1)}$ which is negligible. Thus, the CCA2 adversary's advantage in such cases is also negligible.*

# 5. Conclusions

In this paper, a cryptosystem that is robust against man-in-the-middle attack (active adversary) is proposed. This cryptosystem is a probabilistic one and uses a pre-coding algorithm to encode input messages. We use a secure knapsack-based cryptosystem for performing permutations and random data padding to the message. We showed that the cryptosystem is efficient and secure against active adversary. Another advantage of the proposed cryptosystem is that one can simply implement it in both hardware and software.

# References

[1] Coster, M. J. LaMacchia, B. A. Odlyzko, A. M. and Schnorr, C. P., "An improved low-density subset sum algorithm," in Advances in Cryptology, EUROCRYPT'91, pp. 54–67, LNCS, Vol. 547, 1991.

[2] Dent, A. W.,"A Brief History of Provably-Secure Public-Key Encryption", Progress in Cryptology, AFRICACRYPT 2008, LNCS, Vol. 5023, pp. 357-370.

[3] Lagarias, J. C. and Odlyzko, A. M., "Solving low-density subset sum problems," J. Ass. Comput. Much. vol. 32, no. 1, pp. 229-246. Jan. 1985.

[4] Moret, B. M., Theory of Computation. Addison-Wesley, Reading, 1997.

[5] Menezes, A., Oorschot P. van and Vanstone, S., Handbook of Applied Cryptography. CRC Press, 1997.

[6] Rackoff, C. and Simon, D., "Non-interactive zero-knowledge proof of knowledge and chosen ciphertext attack," In Advances in Cryptology, Crypto '91, LNCS, Vol. 576 pp. 434–444, 1991.

[7] Rasheed, M. F., "Modeling virus propagation in P2P networks," International Journal of Computer Science Issues (IJCSI), Vol. 9, Issue 2, No 2, pp. 580-587, March 2012.

[8] Rastaghi, R. and Dalili Oskouei, H. R., "Cryptanalysis of a public-key cryptosystem using lattice basis reduction algorithms," International Journal of Computer Science Issues (IJCSI), Vol. 9, Issue 5, No 1, pp.110-117, September 2012.

[9] Reidemeister, T. and Böhm, K., "Malicious Behavior in Content-Addressable Peer-to-Peer Networks," IEEE, 3rd Annual Communication Networks and Services Research, pp. 319 – 326, 2005.

[10] Rivest, R. L., Shamir, A. and Adleman, L., "A method for obtaining digital signatures and public key Cryptosystems," Communications of the ACM, vol. 21, pp. 120–126, Feb. 1978.

[11] Shahabadkar, R. and Pujeri, V. R., "Hybrid framework for mitigating illegitimate Peer Nodes in Multimedia file sharing in P2P," International Journal of Computer Science

Issues (IJCSI), Vol. 9, Issue 1, No 2, pp.263-271, January 2012.

[12] Shamir, A., "A Polynomial-time Algorithm for breaking the basic Merkle-Hellman cryptosystem", *Proceedings of the IEEE Symposium on Foundations of Computer Science, New York*, pp. 145-152, 1982.

[13] Merkle, R. C. and Hellman, M. E., "Hiding Information and Signatures in Trapdoor Knapsacks," IEEE Trans. on Information Theory, Vol. 24, pp. 525-530, 1978.

[14] Yang, Y. and Yang, L., " A Survey of Peer-to-Peer Attacks and Counter Attacks," International Conference on Security & Management (SAM 2012), pp. 176-182, 2012.

[15] Zhang, W., Wang B. and Hu Y., " A New Knapsack Public-Key Cryptosystem," IEEE, Fifth International Conference on Information Assurance and Security, pp.53-56, 2009.